\documentclass[twocolumn,aps,prx, showkeys
]{revtex4-2}
\usepackage{graphicx}
\usepackage{subcaption}
\usepackage{dcolumn}
\usepackage{multirow}
\usepackage{physics}
\usepackage{mathtools}
\usepackage{bm}
\usepackage{xcolor}
\usepackage{comment}
\usepackage{natbib}
\usepackage{ragged2e}
\usepackage{blindtext}
\usepackage{lipsum, babel}
\usepackage{amsmath}
\usepackage{makecell} % Allows multi-line headers
\usepackage{array} % Allows custom column widths
\usepackage{soul}
\usepackage[hidelinks]{hyperref}
\AtBeginDocument{}

\begin{document}

\preprint{APS/123-QED}

\title{Distilled remote entanglement between superconducting qubits across optical channels}

\author{\small Nicolas Dirnegger\textsuperscript{1,2}, Moein Malekakhlagh\textsuperscript{1}, Vikesh Siddhu\textsuperscript{1}, 
Ashutosh Rao\textsuperscript{1}, Chi Xiong\textsuperscript{1}, Muir Kumph\textsuperscript{1}, Jason Orcutt\textsuperscript{1}, 
Abram Falk\textsuperscript{1}}
\thanks{\href{mailto:alfalk@us.ibm.com}{alfalk@us.ibm.com}}

\affiliation{\small \textsuperscript{1}IBM Quantum, IBM T.J. Watson Research Center, NY 10598, USA}
\affiliation{\small \textsuperscript{2}Electrical and Computer Engineering Department, University of California, Los Angeles, CA 90095, USA}

\begin{abstract}
A promising quantum computing architecture comprises modules of superconducting quantum processors linked via optical channels using quantum transducers. As quantum transducer hardware improves, a need has arisen to understand the quantitative relationship between transducer-device characteristics and the strength of the resulting remote entanglement. Using Monte Carlo simulations that incorporate 2-to-1 and 3-to-1 entanglement distillation methods, our model maps transducer device performance up to system-level channel performance, thereby allowing the performance of remote entanglement approaches to be compared and optimized. We find the Extreme Photon Loss (EPL) distillation protocol to be particularly high performing. Moreover, even without distillation, present-day transducers with added noise of $N_{add} = 0.5$ photons are at the threshold of enabling remote Bell pairs with fidelities exceeding 50\%. If the next generation of transducers can improve by 3 orders of magnitude in added noise, efficiency, and repetition rates, then they would allow for remote two-qubit gates achieving 99.7\% fidelities at MHz rates. These results set practical targets for transducers to be ready for deployment into modular quantum computing systems.
\end{abstract}

%Use showkeys class option if keyword
                              %display desired

\maketitle
\section*{Introduction} \label{section:Introduction}
\begin{figure*}[ht]
    \centering
    \includegraphics[width=0.9\linewidth]{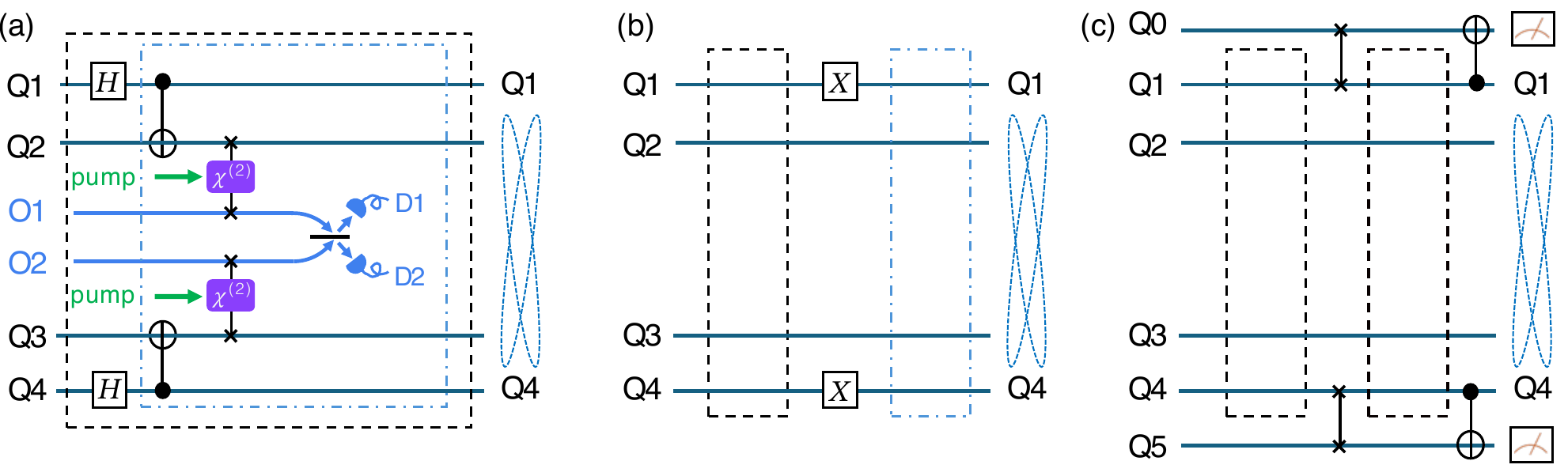}
    \caption{\justifying Circuit diagrams for different remote entanglement procedures. (a) The 1-click protocol using four superconducting qubits (Q1-Q4). Q1 and Q4 are the data qubits that are to be remotely entangled. Q2 and Q3 are the interface qubits that are swapped to optical photon channels O1 and O2 via the quantum transducer (denoted $\chi^{(2)}$). After upconversion, the optical photons are sent to detectors D1 and D2 through a 50:50 beamsplitter. (b) The 2-click protocol comprises two 1-click protocols with with an intervening $\pi$-pulse. The dashed boxes are instances of the 1-click protocol from (a). (c) Remote entanglement with the EPL distillation protocol. A Bell pair is heralded and swapped to memory qubits Q0 and Q5. After creation of second Bell pair, an interference measurement is applied between the memory qubits and data qubits, and a measurement of $\ket{11}$ on the memory qubits heralds successfully distilled entanglement of Q1 and Q4.}
    \label{fig:remote_entanglement}
\end{figure*}
Microwave-to-optical quantum transducers enable coherent conversion between microwave and optical photons \cite{zhong2022microwave,zhao2023electro,mirhosseini2020superconducting,wang2022high,kumar2023quantum,wang2022quantum,weaver2023efficient,sahbaz2024terahertz, Dhara2023PhysRevRes, wu2021deterministic, zhong2024pulsepumped, weaver2025scalable}. They are the key hardware resource needed for microwave-based quantum information systems to be networked over room-temperature connections. In recent years, transducer platforms like electro-optics \cite{wang2022high,zhao2023electro,zhong2022microwave}, optomechanics \cite{jiang2020efficient,simonsen2019sensitive,weaver2024integrated}, and atoms \cite{kumar2023quantum,imamouglu2009cavity,verdu2009strong} have experienced rapid advances. Experiments employing these transducers have recently reached the level of demonstrating entanglement between microwave and optical modes \cite{shi2024entanglement,meesala2024quantum, meesala2024mwoentanglement, meesala2024nonclassical}. 

As transducer hardware improves, there is a need to better understand the relationship between the performance of the transducer as an isolated device and that of the associated quantum channel. Transducer performance is often abstracted to parameters such as the microwave-to-optical conversion efficiency ($\eta$), added noise ($N_{add})$, and bandwidth (BW). Mapping these parameters onto channel-level metrics such as fidelity and the rate of entangled pair generation is key to determining when and how quantum transducers can be deployed.  

Moreover, the achievement of high-fidelity remote entanglement will require mitigation of noise and other errors through techniques such as entanglement distillation \cite{bennett1996purification,chi2012efficient,rozpkedek2018optimizing}. In entanglement distillation, an array of imperfect Bell pairs is transformed into an array of higher fidelity but fewer Bell pairs \cite{nickerson2014freely,beukers2024remote}. Entanglement distillation could be  key to transducer-linked channels with fidelities that match the $90\%+$ fidelity of all-microwave-based long distance gates \cite{wu2024modular,zhao2021practical}. Determining how distillation protocols perform in the context of heralded entanglement via transduction, and which protocols are optimal, are important open questions.

To address these questions, we developed a hybrid quantum/classical Monte Carlo model of remote entanglement that maps transducer device-level metrics to the fidelity and entanglement rate of quantum channels. Our work complements  a previous all-analytical approach \cite{zeuthen2020figures} while providing advantages in certain regimes. First, our model considers multi-photon noise/transduction events, whereas the all-analytical model did not. These multi-photon events are particularly important in the high-$N_{add}$ regime ($N_{add}\approx 1$), where today's state-of-the-art transducers are mostly operating. See Appendix \ref{AppendixB} for a tabulation of device performance levels of current state-of-the-art transducers. Second, the statistical properties of outcome distributions are readily obtained. Third, more complicated entanglement models can be readily considered, including such as those including distillation and the effect of quantum memory relaxation and dephasing.

We use our framework to model single-click heralded entanglement swapping \cite{cabrillo1999creation,campbell2008measurement,narla2016robust,kurokawa2022remote}, and then extend it to the Barrett-Kok protocol \cite{barrett2005efficient} and simple 2-to-1 \cite{bennett1996purification, chi2012efficient, rozpkedek2018optimizing, DEJMPS1996,dur1999quantum} and 3-to-1 (see appendix \cite{chi2012efficient} (and \ref{AppendixC}) entanglement distillation schemes . Recent experimental demonstrations prove the feasibility of implementing these distillation protocols in qubit testbeds (not involving transducers)  \cite{yan2022entanglement,knaut2024entanglement}. 

We conduct our analysis from the standpoint of a red-detuned transduction scheme \cite{tsang2010}, whereby microwave photons originating in superconducting qubits are transferred to microwave transducers and upconverted into optical photons. However, our analysis could be readily adapted to the blue-detuned scheme, where transducers are operated as sources of microwave-optical entangled pairs through spontaneous parametric downconversion \cite{zhong2022microwave}.

Our model generates several practical insights into transduction protocols. We find that the Extreme Photon Loss (EPL) distillation protocol allows for Bell pairs to be generated with fidelities nearly identical to those deriving from the Barrett-Kok protocol, but at rates that scale more favorably with \(\eta\). Looking forward to performance levels required for practical deployment, our model shows that once \(\eta\) and \(N_{add}\) are improved by 2 orders of magnitude each, this protocol can generate Bell pairs with entanglement fidelities exceeding 90\% at a 10 kHz rate. With 3 orders of magnitude of improvement each, 99+\% fidelities can be obtained at 100 kHz rates.

\section*{Methods}
\subsection*{Entanglement swapping} \label{section:RemoteEntanglement}
We adapt a canonical entanglement swapping scheme \cite{dur1999quantum,pan1998experimental} to our system with superconducting qubits and transducers. The superconducting qubits are connected to the quantum transducers through interface qubits, and the transducers themselves are abstracted to being microwave resonators coupled to optical resonators through a nonlinear $\chi^{(2)}$ medium (\autoref{fig:remote_entanglement}(a)). The goal of this protocol is to generate $\ket{\Psi^\pm}$ Bell pairs across an optical channel linking two superconducting qubits:

    \begin{equation}
    \ket{\Psi^\pm}=\frac{1}{\sqrt{2}}(\ket{ge}\pm\ket{eg}),
    \end{equation}

\noindent where $\ket{e}$ and $\ket{g}$ are the qubit ground and excited states, respectively.

\begin{enumerate}
    \item Two pairs of superconducting qubits are initialized into their ground states. Q1 and Q4 are the two data qubits in two remote dilution refrigerators that are to be entangled, and Q2 and Q3 are interface qubits, which connect the qubits to the transducers. Alice and Bob each have a data qubit, interface qubit, and optical photon number in their transducer's optical resonator: \(\ket{\Psi}_{A,B}=\ket{\Psi_{\text{data}}{\Psi_{\text{interface}}{\Psi}_{\text{optical}}}}_{A,B}\). Q1 and Q4 are put into pairwise superposition states via  Hadamard gates, resulting in:

    \begin{equation}
    \ket{\Psi}_{A,B} = \frac{1}{\sqrt{2}} (|g\rangle + |e\rangle )\ket{g0}
    \end{equation}

    where $\ket{0}$ refers to the Fock state of the transducer's optical resonator.
    \item Pairwise CNOT gates then entangle the data qubits (Q1 and Q4) with the interface qubits (Q2 and Q3):
    \begin{equation}
    \ket{\Psi}_{A,B} = \frac{1}{2}(\ket{gg}+\ket{ee})\ket{0}
    \end{equation}
    \item The interface qubit states are then swapped to the transducers' unoccupied microwave resonators. At this point, the transducers' pumps are pulsed, and in the limit of \(\eta=1\) (100\% transduction success), the states in these microwave resonators are then swapped to previously unoccupied optical resonators, resulting in:
    \begin{equation}
    \ket{\Psi}_{A,B} = \frac{1}{2}(\ket{gg0}+\ket{eg1}).
    \end{equation}
    \item The role of the interface qubits is now complete, so we no longer include them in our notation. Alice and Bob's data qubits and optical resonator states are:
    \begin{equation}
    \ket{\Psi}_{A}\ket{\Psi}_{B} = \frac{1}{2}(\ket{g0}+\ket{e1})_A(\ket{g0}+\ket{e1})_B
    \end{equation}
    \item The photons are then directed to the 50:50 optical beamsplitter, whose action is:

    % Beamsplitter matrix if that is preferred
    % \begin{equation}
    % \begin{pmatrix}
    % \ket{00}\\\ket{01}\\\ket{10}\\\ket{11}
    % \end{pmatrix}_{ab}\rightarrow
    % \frac{1}{\sqrt{2}}\begin{pmatrix}
    % \sqrt{2} & 0 & 0 & 0 & 0 & 0 \\
    % 0 & 1 & -1 & 0 & 0 & 0 \\
    % 0 & 1 & 1 & 0 & 0 & 0 \\
    % 0 & 0 & 0 & 0 & 1 & -1
    % \end{pmatrix}
    % \begin{pmatrix}
    % \ket{00}\\\ket{01}\\\ket{10}\\\ket{11}\\\ket{02}\\\ket{20}
    % \end{pmatrix}_{cd}
    % \end{equation}
    
    \begin{equation} 
    \ket{00}_{ab}\rightarrow\ket{00}_{cd} 
    \end{equation}
    \begin{equation}
    \ket{10}_{ab}\rightarrow\frac{1}{\sqrt{2}}(\ket{10_{cd}}+\ket{01}_{cd}) 
    \end{equation}
    \begin{equation}
    \ket{01}_{ab}\rightarrow\frac{1}{\sqrt{2}}(\ket{10}_{cd}-\ket{01}_{cd}) 
    \end{equation}
    \begin{equation}
    \ket{11}_{ab}\rightarrow\frac{1}{\sqrt{2}}(\ket{20}_{cd}-\ket{02}_{cd})
    \end{equation}

    where $a$ and $b$ are the beamsplitter's input ports, and $c$ and $d$ are its output ports. The state of the two qubits plus the two photon states after the beamsplitter is then:
    \begin{equation}
    \begin{split}
    \\\ket{\Psi}_{A,B}=\ket{gg}\frac{\ket{00}_{cd}}{2}+\ket{ge}\frac{\ket{01}_{cd}-\ket{10}_{cd}}{\sqrt{2}}
    \\ 
    +\ket{eg}\frac{\ket{01}_{cd}+\ket{10}_{cd}} {\sqrt{2}}+\ket{ee}\frac{\ket{20}_{cd}-\ket{02}_{cd}}{2} 
    \end{split}
    \label{eq:keyState}
    \end{equation}
    \item Thus, the detection of exactly one photon in one of the detectors (i.e. in the $c$ or $d$ optical channel) projects Q1 and Q4 into one of the $\ket{\Psi^\pm}$ Bell states, where the $\pm$ phase depends on which detector clicked.

\end{enumerate}

Optical single photon detectors will often not be photon-number resolving, i.e. they will "click" once whether they receive one or more photons. As a result of this, as well as sub-unity $\eta$, double excitation errors can arise, in which $\ket{ee}$ states are heralded instead of the desired $\ket{\Psi^\pm}$ states.
A well-known approach to mitigate these errors is the Barrett-Kok protocol \cite{barrett2005efficient,narla2016robust}, otherwise known as the 2-click protocol. Here, after an optical photon is detected, pairwise $\pi$ pulses are applied to the data qubits, and transduction and heralding are attempted again \autoref{fig:remote_entanglement}(b). The $\pi$ pulses rotate $\ket{ee}$ states into $\ket{gg}$ states, which will not produce optical photons in the second heralding round, thereby excluding them from degrading the heralded fidelity.

The 2-click protocol also transforms local phase errors deriving from unequal lengths of the setup's optical arms to a global phase \cite{barrett2005efficient}, thereby mitigating the need for active path-length stabilization during transduction. However, the 2-click protocol requires two exactly sequential successful transduction events, leading to very low entanglement rates in the $\eta\ll1$ regime \cite{zeuthen2020figures,rozpkedek2018optimizing,beukers2024remote}. 

\subsection*{Entanglement Distillation}
Entanglement distillation corrects errors in noisy Bell states through processes of interference and post-selection. Different distillation protocols address different types of noise and loss mechanisms.

The extreme photon loss (EPL) protocol \cite{bennett1996purification, nickerson2014freely} is specifically designed to handle highly lossy quantum communication networks. When applied to $\ket{\Psi^\pm}$ states, it corrects generalized amplitude damping errors to both $\ket{gg}$ and $\ket{ee}$ states. Like the 2-click protocol, it also converts local phase errors deriving from path length fluctuations to a global phase that can be ignored. 

In the EPL protocol, a single Bell pair is first heralded (\autoref{fig:remote_entanglement}(c)). The Bell pair is swapped to a pair of memory qubits (Q0 and Q5), and a second Bell pair is then heralded in the data qubits (Q1 and Q4). Pairwise CNOT gates interfere the two Bell pairs and the state of the memory qubits is measured. If $\ket{Q2,Q3}=\ket{1,1}$, then the sequence is successful and the entanglement between Q1 and Q4 is accepted. Otherwise, the qubits are reinitialized, and the sequence is repeated.  
% An outline the protocol is described below and shown in ~\autoref{fig:remote_entanglement}(c).

% \begin{enumerate}
%     \item A first Bell pair is heralded, which can have a phase error $(\phi)$ and a double excitation error component:($|\Psi^\phi \rangle = \frac{1}{\sqrt{2}} (|01\rangle + e^{i\phi} |10\rangle$) to store entanglement resource after distillation.

%     % \[\Psi = ((1-r)^2|\Psi^\phi \rangle |\Psi^\phi \rangle + (1-r)r|\Psi^\phi \rangle |11\rangle\]
%     % \[+ r(1-r)|11\rangle |\Psi^\phi \rangle + r^2 |11\rangle |11\rangle)\]
    
%     \item Alice and Bob both apply a CNOT gate on their qubits, choosing control and target qubits respecitvely.
    
%     \item One of the qubit pairs is to be measured in the z basis. Any result other than “11” is rejected. Only the first of the four possibilities listed above satisfies this condition. We collect terms, measure the target qubits (denoted in bold) and obtain:
    
%     % \[(|01\rangle (\boldsymbol{|00\rangle} + e^{i\phi} \boldsymbol{|11\rangle}) + e^{i\phi}|10\rangle (\boldsymbol{|11\rangle} + e^{i\phi} \boldsymbol{|00\rangle}))\]
%     % \[= (|01\rangle + e^{2i\phi}|10\rangle) \boldsymbol{|00\rangle} + e^{i\phi}(|01\rangle + |10\rangle) \boldsymbol{|11\rangle} \]
    
%     \item If the measurement results of Alice and Bob coincide, the distillation is successful, and the qubit pair is stored as state; if the measurement results fail, they claim the distillation failed and the qubit pair will be discarded. 
% \end{enumerate}

Like many other entanglement distillation methods, the EPL protocol requires input fidelities of $F>0.5$ to be effective, as well as ebit generation rates exceeding memory decay rates.

The EPL protocol contrasts with other well-known bipartite distillation protocols such as the BBPSSW protocol \cite{bennett1996purification} and the DEJMPS protocol \cite{DEJMPS1996}, which do not correct both $\ket{gg}$ and $\ket{ee}$ errors. We also model the Chi 3-to-1 protocol \cite{chi2012efficient}. In general, the choice of entanglement distillation protocol involves identifying dominant error sources and deciding how to trade off fidelity improvement, entanglement rates, and resource demands.

The core of our Monte Carlo approach is a semiclassical generalized amplitude damping and dephasing model of entanglement swapping in the presence of added noise. When one of the optical photon detectors ``clicks," a Bell pair may be heralded, but errors deriving from noise photons and double excitations may also cause undesired states to be heralded. Our model captures these possibilities by treating qubit excitation and transduction events as random variables following classical statistical processes.

We follow the circuit in \autoref{fig:remote_entanglement}(a), except that the Hadamard gates are generalized to single-qubit gates that excite qubits Q1 and Q4 into pairwise superposition states \(\sqrt{1-p_e}\ket{g}+\sqrt{p_e}\ket{e}\), where the parameter $p_e$ specifies the qubit's probability of excitation \cite{siddhu2023optimal,pal2018qudit}. We model qubit states in the measurement basis with Bernoulli random variables \(x_{1}\) and \(x_{2}\), which can be 0 or 1 with \(\langle x_{1}\rangle=\langle x_{2}\rangle=p_{e}\), representing the excitation level of Q1 and Q4 respectively. These two qubits are then pairwise entangled with the two superconducting interface qubits with a pair of CNOT gates, which sets Q2 to equal Q1 and Q3 to equal Q4.

The states of the interface qubits Q3 and Q4 are then transferred to the microwave resonators of the two transducers. Whether transduction is successful is modeled with two more Bernoulli random variables \(\eta_{1}\) and \(\eta_{2}\) satisfying \(\langle \eta_{j}\rangle=\eta\). Here, $\eta$ is the end-to-end efficiency of the microwave-to-optical channel, not just the efficiency of the transducer. This efficiency is the product of the efficiency of the transducer, the transducer-to-fiber couplers, the pump filters, and the optical photon detectors.

Next, added noise is represented by a discrete number of added noise photons that can populate the microwave resonators and probabilistically get transduced. The number of added noise photons are modeled as whole-number random variables \(n_{1}\)  and \(n_{2}\)  drawn from Poissonian distributions with expectation values \(\langle n_{1}\rangle=\langle n_{2}\rangle=N_{add}\). Like the photons that were transferred to the microwave resonators from the qubits, these noise photons are then probabilistically upconverted to optical frequencies, with efficiencies (\(\eta_{3}\), \(\eta_{4}\), ...), each of which is a Bernoulli random variable like \(\eta_{1}\) and \(\eta_{2}\). At the end of each trial of this simulation, between 0 and 2 qubit-derived photons and between 0 and \(n_{1}+n_{2}\) noise photons have been upconverted and detected. We then classify the density matrix of the data qubit pair as follows: (\autoref{table:monte}):

\begin{enumerate}
    \item \textbf{Successful entanglement swapping}
\\In a successful entanglement swapping sequence (\autoref{table:monte}(a)), exactly one of the data qubits will have been excited, and that excitation from an interface qubit will have been swapped to the transducer, upconverted, and detected. Meanwhile, noise photons will have either not been excited, or they will have been excited but not transduced. In this case, the resulting two-qubit state of the data qubits is one of the two ${\Psi^{\pm}}$ Bell states.

The phase of the Bell state (e.g. whether \(\Psi^{+}\) or \(\Psi^{-}\) is heralded) depends upon which single-photon detector clicks. However, if we define \(\Psi^{+}\) to be the phase if detector 1 clicks, then if detector 2 clicks, a single qubit rotation could be used to rotate \(\Psi^{-}\) into \(\Psi^{+}\) . Thus, for the purposes of this simulation, either detector clicking heralds a successfully generated Bell pair.

    \item \textbf{Amplitude damping to the \(\ket{gg}\) ground state }
\\If neither data qubit was excited, but a detector receives a photon anyway (i.e. it receives a noise photon), then the \(\ket{gg}\) state is heralded instead of the Bell state. This error can be visualized as an amplitude damping of the Bell state to \(\ket{gg}\) (\autoref{table:monte}(b)) .

    \item \textbf{Generalized amplitude damping to the \(\ket{ee}\) state }
\\If both data qubits were excited, and one of those photons or a noise photon has reached the detector, then the \(\ket{ee}\) state is heralded instead of the Bell state. This double excitation error can be visualized as an amplitude damping of the Bell state to \(\ket{ee}\) (\autoref{table:monte}(c)).

    \item \textbf{Phase damping to a mixed state}
\\If exactly one of the data qubits is excited, as is targeted, but a noise photon reaches the detector, then we set the heralded density matrix to the mixed state comprising equal populations of \(\ket{ge}\) and \(\ket{eg}\), as shown in (\autoref{table:monte}(d)). The justification for this density matrix is that noise photons populating the microwave resonator have no phase relationship with the data qubits Q1 and Q4. Therefore, when a noise photon is detected, no information about the relative phases of the two qubits is learned, just as would be the case for a true dark count deriving from the optical detector \cite{zeuthen2020figures}.

    \item \textbf{No photon is detected}
\\When no photon is detected, heralding fails and another attempt at entanglement swapping is performed.

\end{enumerate}

\begin{table} [h]
\renewcommand{\thetable}{\arabic{table}}
\centering
\includegraphics[width=\linewidth]{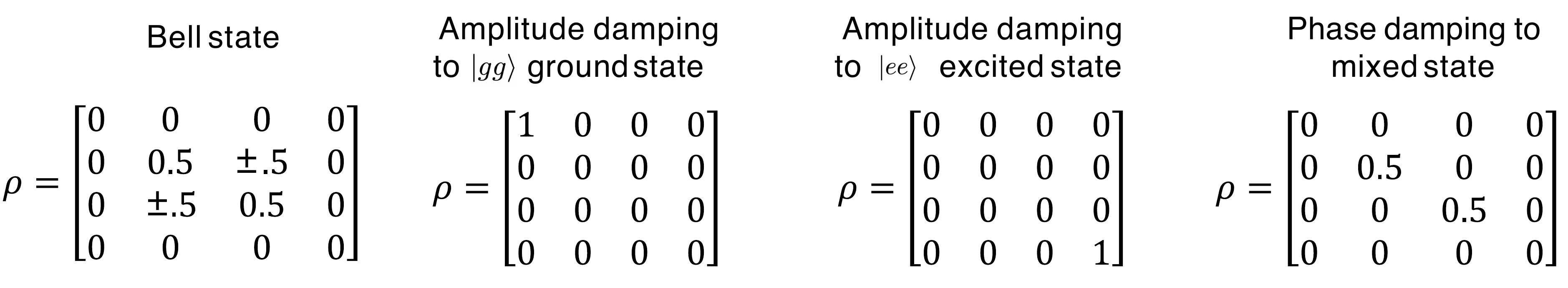}
\caption{\justifying Bloch sphere representations of the 4 possible outcomes of a trial of our Monte-Carlo model, along with the associated density matrices. These four outcomes are: 1), the Bell state $|\Psi^\pm\rangle$, 2) amplitude damping channel to the $|gg\rangle$ state deriving from transduced added noise photons, 3) generalized amplitude damping to the excited state $|ee\rangle$ due to double excitation errors, 4) phase damping channel to the \(\rho_{mixed} =\frac{1}{2}(\ket{ge}\bra{ge}+\ket{eg}\bra{eg}\)) state. This final process represents loss of coherence from the system without a loss of energy.}
\label{table:monte}
\end{table}

After each heralding event, the fidelity of the resulting density matrix to the target  state is calculated. The fidelity is calculated by using 

\begin{equation}\label{Fidelity}
    F_{ic} = Tr[\rho^{'} |\Psi^+\rangle \langle \Psi^+|]
\end{equation}

\noindent where $\rho'_{}$ represents the density matrix after rotating into $|\Psi^+ \rangle$.

Many trials are run, and the fidelity is averaged across these trials. The entanglement rate is calculated by multiplying the attempt frequency by the Bell pair success fraction. For the 2-click and distillation protocols, additional heralding attempts are made, as prescribed by those protocols. Optimal protocols are then identified by adjusting the several parameters that can be used to trade off fidelity and entanglement rate: 
\begin{itemize}
    \item The protocol type (e.g. 1-click, 2-click, EPL, or another distillation protocol).
    \item The parameter \(p_e\).
    \item The transducer's pump laser power or pulse time, which affect both \(\eta\) and \(N_{add}\).
\end{itemize}

The assumptions made in this simulation are as follows:

\begin{enumerate}

    \item The superconducting qubit gates are ideal.
    \item Superconducting qubit measurement fidelities are ideal.
    \item The qubits' relaxation and decoherence are disregarded (except when memory \(T_1\) loss is explicitly included in \autoref{fig:transducerComparison}).
    \item The heralded states correspond to the density matrices as shown in ~\autoref{table:monte}. In particular, the added noise process happens independently of the process of transduction of photons deriving from the superconducting qubits.
    \item The probability of noise photons populating the microwave resonators is a function of optical pump power but independent of other noise photons. As such, the noise photon population probability follows Poissonian statistics.
    \item The detector dark count rates are negligible compared to the rate of upconverted noise photons. This approximation derives from the fact that with a pulsed pump (e.g. tens of nanoseconds), the dark count rate of state-of-the-art superconducting nanowire single-photon detectors is near zero.
    \item Parametric amplification effects are neglected, i.e. the transducers here are functioning solely as microwave-to-optical upconversion devices.
    \item The product of the transducer's pulse time and its bandwidth is 1, leading to an optimal amount of introduced added noise, given the specification of the transducer.
    \item The single photon detectors are not photon-number resolving, i.e. they click if one or more photons reaches them within a time bin.
    \item Either one of the two single-photon detectors can result in a successful Bell state heralding event. If the $\ket{\Psi^+}$ state is being targeted, and the wrong detector clicks, we assume a fast feedforward rotation is used to rotate $\ket{\Psi^-}$ into $\ket{\Psi^+}$.
    \item The microwave resonator of the quantum transducer is narrowband. As a result, the resonator is spectrally selective, and the original spectrum of the noise photons only impacts the rate at which they populate the resonator. 
    
\end{enumerate}

\begin{figure}[h]
    \centering
    \includegraphics[width=\linewidth]{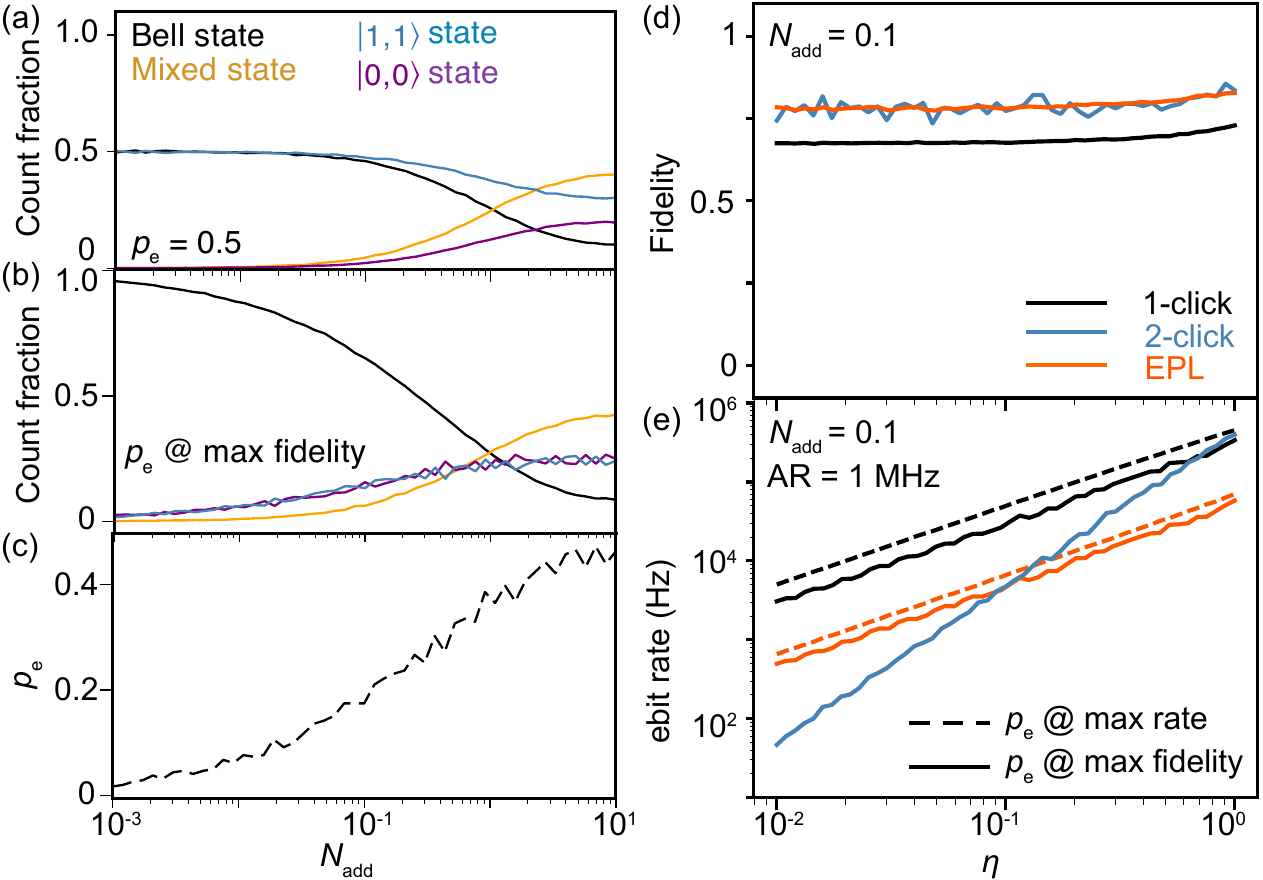}
    \caption{\justifying The fraction of heralded events corresponding to the scenarios shown in Table \autoref{table:monte}  at (a) $p_e=0.5$ and (b) the $p_e$ value that maximizes fidelity of the 1-click protocol. (c) The $p_e$ value that maximizes the fidelity of the 1-click protocol. (d) The fidelity and (e) entanglement rate of the 1-click, 2-click and EPL protocols as a function of \(\eta\), with \(N_{add}\) = 0.1 and the attempt rate = 1 MHz.}
    \label{fig:etaSweep}
\end{figure}

\section*{Results}

In executing our Monte Carlo simulation, we generally average 5000 trials for each set of parameters. 

We start by quantifying the count rate of the noise-induced errors during the 1-click remote entanglement process (\autoref{fig:etaSweep}(a,b)). The fraction of counts resulting in the desired Bell pair state increases as $N_{add}$ decreases, but this fraction saturates at 0.5 if $p_e$ is held constant. This saturation derives from double excitation errors. At \(p_e=0.5\), the entanglement rate for the 1-click and 2-click protocols is maximized, as is the fidelity of the 2-click protocol. 

However, the \(p_e\) value that maximizes fidelity for the 1-click rate depends on \(N_{add}\) (\autoref{fig:etaSweep}(c)). When \(N_{add}\) is high, the optimal \(p_e\) value is high as well, so that the qubit excitation rate can compete with the noise photon rate. As \(N_{add}\) is decreased, the optimal \(p_e\) decreases as well, to minimize double excitation errors.  

The simulated fidelity and entanglement rate show several clear trends as \(\eta\) is swept. The fidelity is fairly insensitive to \(\eta\) (\autoref{fig:etaSweep}(d)). The 2-click and EPL fidelities exceed the 1-click fidelity and are nearly identical to each other. Naturally, the entanglement rate is highly sensitive to \(\eta\). The 2-click entanglement rate scales quadratically with \(\eta\), whereas the entanglement rate for the 1-click and EPL protocols scales linearly with \(\eta\) (\autoref{fig:etaSweep}(e)).

This difference in scaling between the 2-click protocol and the EPL protocol demonstrates the advantage of distillation. This difference derives from the use of the memory in distillation. With the 2-click protocol, two successful transduction events must happen in a row -- hence the \(\eta^2 \) scaling. With the EPL protocol, the second transduction cycle can be buffered \cite{zang2023entanglement, davies2024entanglement,inesta2025entanglement}, i.e. attempted many times in a row, within the limits of the memory's \(T_1\) and \(T_2\) times. Accordingly, a linear scaling with \(\eta\) is achieved. The EPL protocol's benefit is that it allows for a 2-click-like fidelity with a 1-click-like entanglement rate-rate scaling.

\begin{figure}[h]
    \centering
    \includegraphics[width=\linewidth]{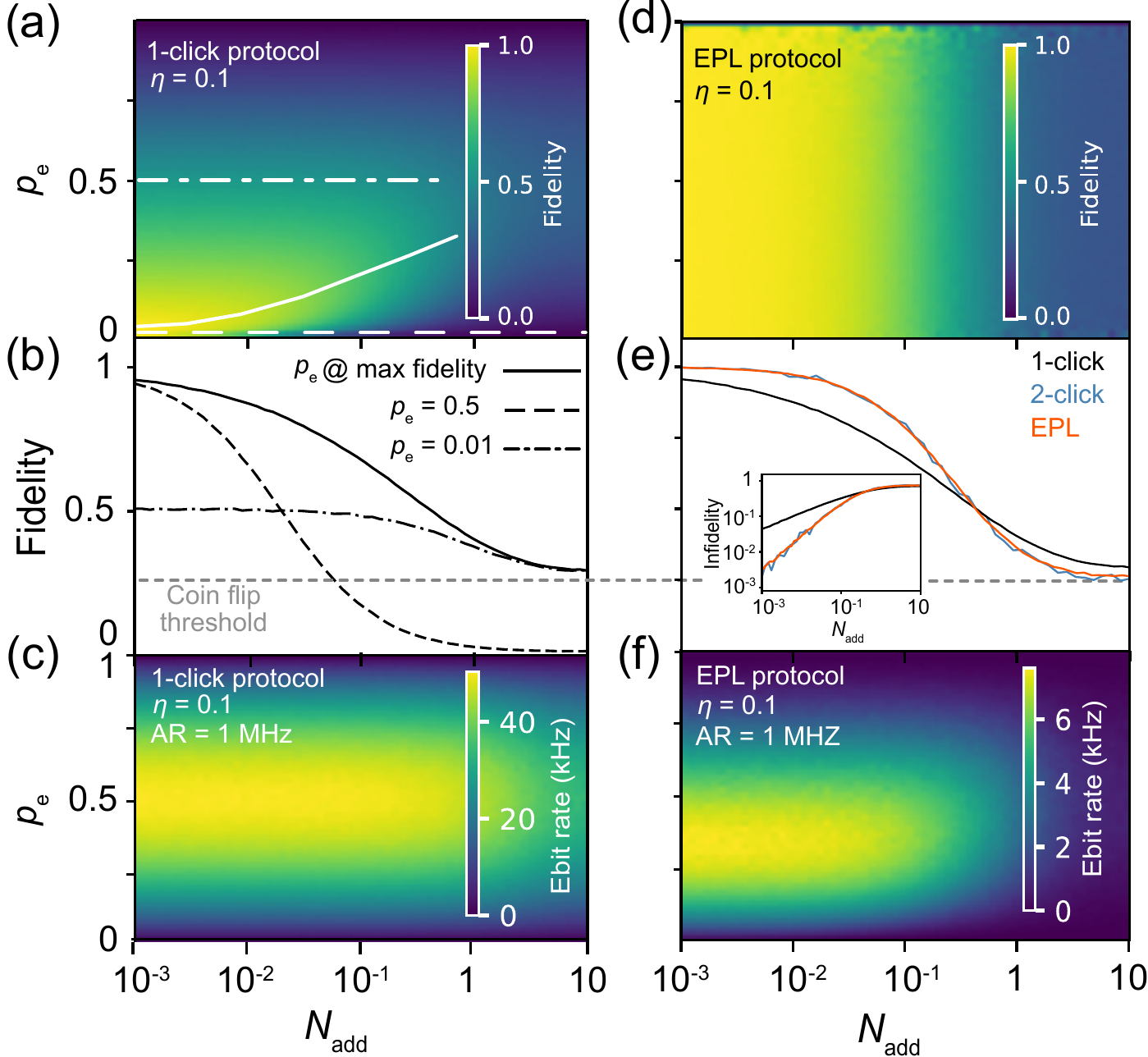}
    \caption{\justifying The fidelity and e-bit rate as a function of \(N_{add}\). (a) The fidelity of the 1-click protocol as a function of \(N_{add}\) and \(p_e\), with \(\eta\) = 0.1. The solid white line corresponds to maximum fidelity line cut. (b) Fidelity vs. \(N_{add}\) line-cuts of (a) at maximum fidelity (solid line), maximum e-bit rate (dashed-dotted line), and minimal near-0 values of \(p_e\) (dashed line). (c) E-bit rate as a function \(N_{add}\) and \(p_e\), with the attempt rate (AR) = 1 MHz. (d) Fidelity of the EPL protocol as a function of \(N_{add}\) and \(p_e\), with \(\eta\) = 0.1. (e) Fidelity and infidelity of the 1-click, 2-click, and EPL protocols as a function of \(N_{add}\) , with \(p_e\) chosen to maximize fidelity for the 1-click protocol. (f) Entanglement rate of the EPL protocol, with AR = 1 MHz, as before.}
    \label{fig:NaddDependence}
\end{figure}

Fidelity is a strong function of \(N_{add}\) (\autoref{fig:NaddDependence}(a-c)). At very high \(N_{add}\), the two-qubit fidelity approaches 0.25, the fidelity that is expected if each qubit's state derives from a coin flip. As \(N_{add}\) is decreased, the optimized 1-click fidelity reaches 90\% at \(N_{add}=10^{-2}\). However, the entanglement rate for the fidelity-optimized 1-click protocol drops precipitously as \(N_{add}\) is lowered, because the optimal \(p_e\) value decreases with \(N_{add}\) (\autoref{fig:NaddDependence}(c)).

On the other hand, the EPL fidelity is insensitive to \(p_e\) (\autoref{fig:NaddDependence}(d)). Moreover, the fidelities for both the 2-click protocol and the EPL distillation protocols improve much faster than the 1-click fidelity as \(N_{add}\) is lowered (\autoref{fig:NaddDependence}(e). They reach 0.997 at \(N_{add}\) = \(10^{-3}\), corresponding to over an order of magnitude improvement (in infidelity) over the 1-click protocol. The EPL entanglement rate is maximized at \(p_e\) =  0.3 (\autoref{fig:NaddDependence}(f). The entanglement rate for all three protocols is insensitive to \(N_{add}\) until \(N_{add}\) exceeds a threshold. This threshold is \(N_{add}\sim1\) for the 1-click protocol and \(N_{add}\sim0.1\) for the 2-click and EPL protocols (\autoref{fig:NaddDependence}(c,f).

\begin{figure}[h]
    \centering
    \includegraphics[width=0.7\linewidth]{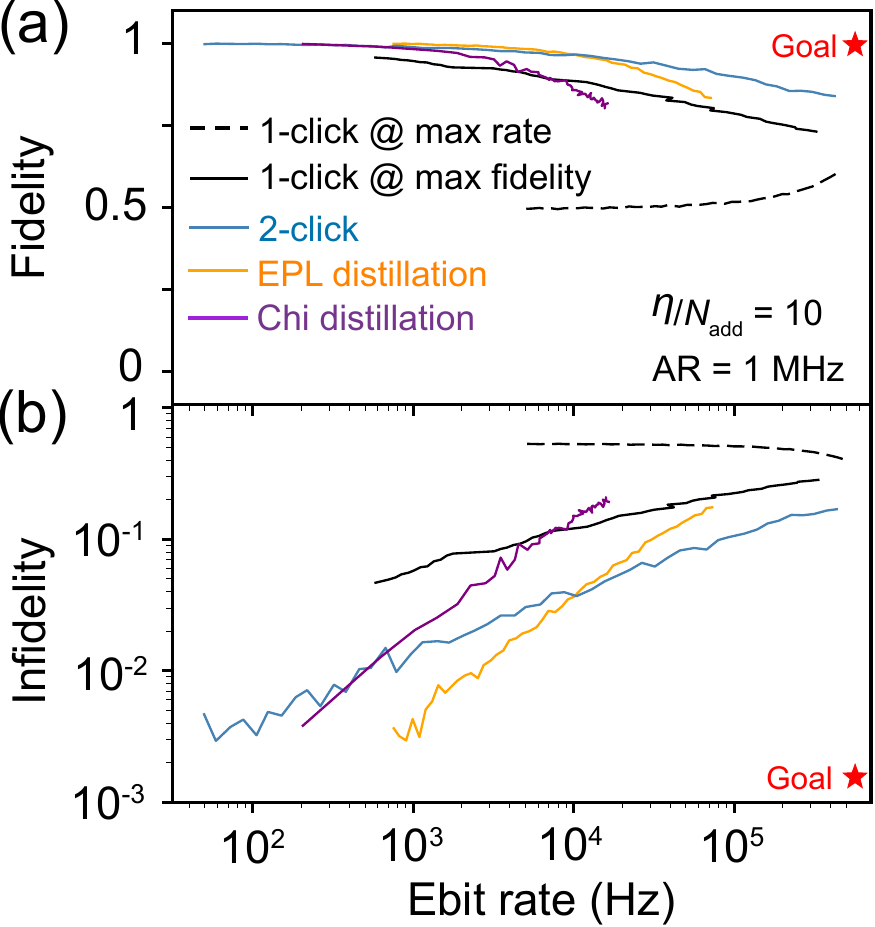}
    \caption{\justifying A comparison of e-bit rate vs. (a) fidelity and (b) infidelity for the 1-click, 2-click, EPL 2-to-1 distillation, and Chi 3-to-1 distillation protocols. To generate these curves, we vary \(N_{add}\) (from \(10^{-3}\) to \(10^{-1}\)) and \(\eta\) (from \(10^{-2}\) to 1) together, while fixing the ratio \(\eta / N_{add} = 10\), and setting the attempt rate to 1 MHz. For the 1-click protocol, \(p_{e}\) can be chosen to maximize either fidelity or entanglement rate.}
    \label{fig:FidelityVsER}
\end{figure}

To understand how these variables interrelate, we consider a scenario in which both \(\eta\) and \(N_{add}\) are directly proportional to the laser pump power and then vary that power while fixing their ratio to \(\eta / N_{add} = 10\). This ratio, about one order of magnitude better than the performance of current state-of-the-art transducers, was selected as it is a near-deployment-ready regime. We can then plot fidelity vs. entanglement rate and identify outperforming protocols (\autoref{fig:FidelityVsER}). We find that for low entanglement rates (below 10 kHz), the EPL protocol outperforms the 2-click protocol, whereas the 2-click protocol outperforms in the high \(\eta\) regime when the entanglement rate is above 10 kHz. We also plotted the performance of the Chi 3-to-1 distillation protocol \cite{chi2012efficient} here, but it was not found to be optimal in any regime here.

We have not yet considered the effect of \(T_{1}\) or \(T_{2}\) decay of the first Bell pair while waiting for the second Bell pair to be heralded\cite{siddhu2024dampingdephasing,aliferis2009faulttolerant}. With present-day transducers and superconducting qubits as memories, this decay is a significant consideration. Accordingly, we also modeled distillation in the presence of qubit relaxation. To be realistic, we started with present-day performance parameters given by the current generation of optomechanical transducers \cite{jiang2020efficient,mirhosseini2020superconducting,han2021microwave,andrews2014bidirectional,fang2016optical} and specified \(\eta=10^{-4}\) ,  \(N_{add}=0.5\), and a repetition rate of 100 kHz. We modeled the superconducting qubits as having $T_{1} = 300$ $\mu s$, which is typical of high-performing transmons ~\cite{place2021new,bal2024systematic}. We find that distillation is not possible with present-day transducers, unless a quantum memory is used that has a significantly longer lifetime than that of a superconducting qubit.

\begin{figure}[h]
    \centering
    \includegraphics[width=\linewidth]{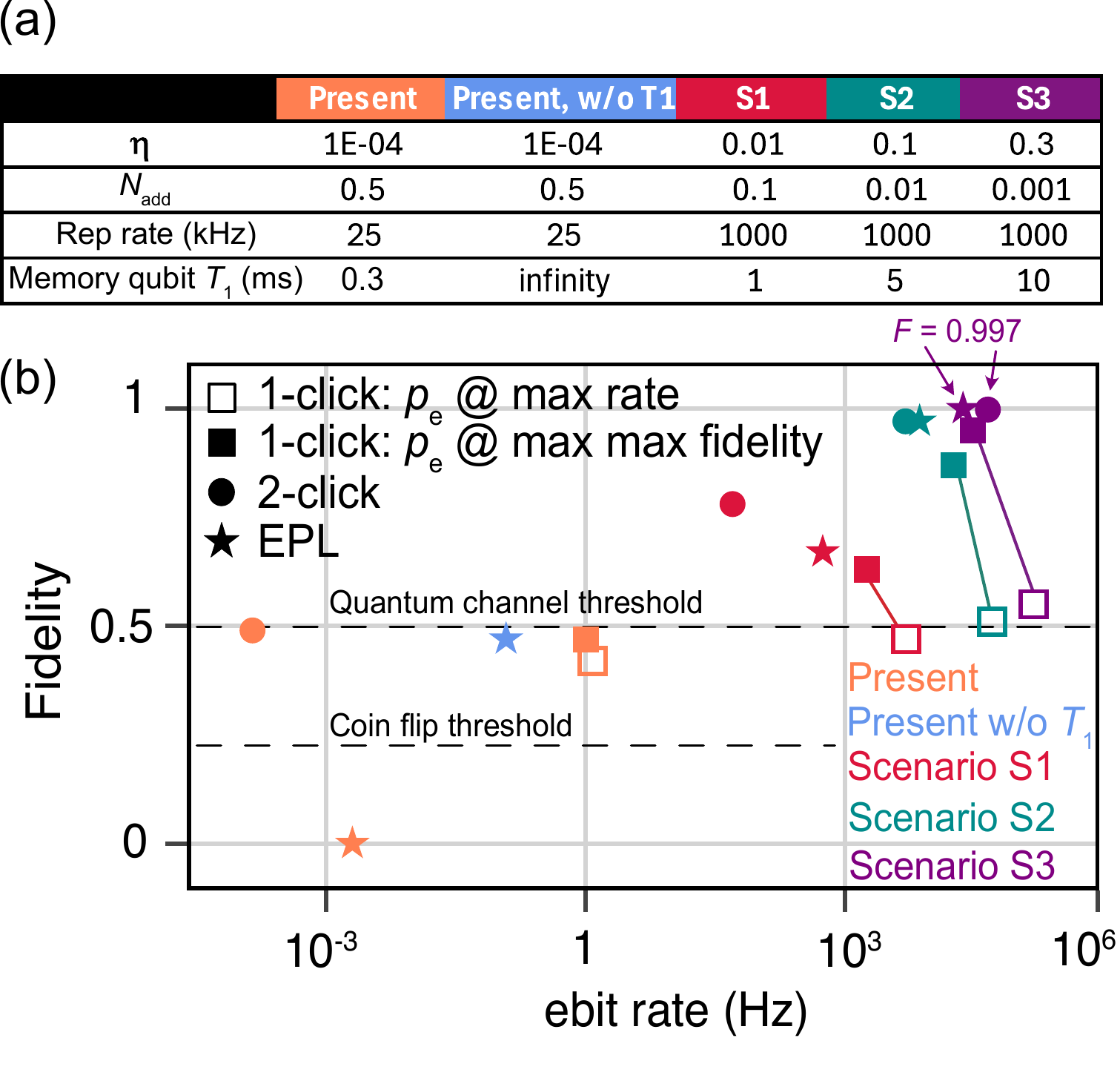}
    \caption{\justifying A comparison of entanglement rate and fidelity for the 1-click, 2-click, and EPL protocols under five different scenarios. "Present" is the approximate state-of-the-art. "Present w/o \(T_1\)" is the same specifications except the memory qubit's \(T_1\) time is set to be infinity. For this scenario, only the EPL protocol's metrics are shown, as the others do not use a memory. S1-S3 are three improvement scenarios. The solid lines connecting the open and closed squares represent points on the fidelity vs. entanglement rate that can be achieved by the 1-click protocol through the choice of \(p_e\). }
    \label{fig:transducerComparison}
\end{figure}

However, the 1-click protocol comprising devices performing at these levels should be at the threshold of being able to demonstrate remote entanglement above 50\% (\autoref{fig:transducerComparison}). We also considered three improvement scenarios (S1, S2, and S3) and observe that as the transducer and \(T_{1}\) performance metrics improve, distillation becomes viable. In improvement scenario S2 (\autoref{fig:transducerComparison}), achieving a distilled fidelity of 0.97 at an entanglement rate of 7.5 kHz requires a 1000\(\times\) increase in \(\eta\), and 50\(\times\) reduction to \(N_{add}\), as well as repetition rate improvements. In scenario S3, consisting of \(\eta=0.3\), \(N_{add}=0.001\), \(T_{1} = 10\) ms, and a repetition rate of 1 MHz, a Bell-pair fidelity of 0.997 and entanglement rates exceeding 100 kHz are achievable. These results underscore the critical role of reducing noise and increasing transducer efficiency \cite{rozpkedek2018optimizing}, along with reducing other insertion losses in the optical path.

\hfill

\section*{Discussion}

Besides the simple distillation schemes considered here, there are other distillation schemes that could be implemented. Other 2-to-1 distillation protocols \cite{zhao2021practical,yan2022entanglement} could be considered, or the distillation schemes considered here could be applied recursively. More sophisticated distillation protocols include hashing protocols, breeding protocols, stabilizer protocols, and many others \cite{Dur2007, rozpkedek2018optimizing}. Furthermore, the output of Bell pairs generated by the 2-click protocol could be used as input states to any of these distillation schemes. One issue that may need to be addressed in the future is that the sequential attempts of probabilistic entanglement generation will always in practice lead to non-identical input states that impose limitations to the performance of entanglement distillation \cite{zang2025no,zang2025entanglement}. In general, though, selecting the appropriate distillation protocol will involve a decision regarding how to trade off fidelity and entanglement rate, and the EPL protocol already performs well with a low overhead.

Another question is when multiple transducers per qubit should be considered \cite{weaver2025scalable}. This would certainly help with the entanglement rate and/or enable more advanced distillation protocols than simple 2-to-1 protocols. Because every transducer comes with a power, space, and wiring overhead, this approach could be impractical if a high optical link density is required. Nonetheless, it should be considered if the number of required system-to-system links is low.

Our analysis shows that present-day transducers are nearing the threshold of being capable of distributing Bell states with fidelities of 50\%, a milestone for remote entanglement generation \cite{zhong2022microwave,kurokawa2022remote}. The EPL  protocol is a particularly promising distillation protocol, correcting both $\ket{gg}$ and $\ket{ee}$ errors and providing first-order protection against phase fluctuations caused by path-length differences \cite{nickerson2014freely}. Its fidelity matches that of the 2-click protocol across a wide parameter range, yet its entanglement rate scales linearly with \(\eta\). 

For transducers to rival the performance of current all-microwave links with fidelities exceeding 90\% \cite{wu2024modular,zhao2021practical,zhong2021deterministic}, significant advances to transducer hardware are still needed. Distillation with today's transducers and quantum memories comprising today's superconducting qubits remains infeasible, as the memory dephasing rate exceeds the entanglement rate. 

Nonetheless, the application of entanglement distillation to remote entanglement of superconducting qubits has a clear pathway forward. As transducer device progress is realized, the improvement scenarios we considered show that high fidelities are possible. Moreover, quantum memories such as the silicon-vacancy color center could mitigate decoherence and relaxation of the quantum memory  \cite{knaut2024entanglement}. Additionally, the use of hybrid quantum architectures that combine solid-state memories with photonic systems \cite{kurokawa2022remote,yan2022entanglement} could address current limitations in transducer efficiency and noise. Finally, beyond the simple distillation protocols that we considered here, more sophisticated distillation protocols could be tailored to transducer noise characteristics. Together, these advances could provide a robust framework for long-distance networking of superconducting qubits across across room-temperature optical channels.

\section*{Author Contributions}
The simulations were performed by N. Dirnegger and A. Falk. M. Malekakhlagh, A. Rao, C. Xiong, and V. Siddhu contributed theoretical underpinnings of the model. All authors contributed to the analysis of the results and the writing of the manuscript.

\section*{Competing Interest}
All authors declare no financial or non-financial competing interests.
\begin{acknowledgements}
We acknowledge helpful discussions with Joseph Peetz, Jimmy Ying, Chandni Nagda, and Luke Govia.
\end{acknowledgements}

\section*{Data and Code Availability}
\vspace{-1em}
The data and code that support the findings of this study are available from the corresponding authors upon reasonable request.

\onecolumngrid
\appendix
% \begin{equation}
% \ket{\Psi^+}=\frac{1}{\sqrt{2}}(\ket{10}+\ket{01})
% \end{equation}

% \begin{equation}
% \ket{\Psi^+}\ket{\Psi^+}\rightarrow\textbf{bilocal CNOTs}\rightarrow\ket{\Psi^+}\ket{\Psi^+}
% \end{equation}
% \begin{equation}
% \ket{\Psi^+}\ket{\Psi^+}= \frac{1}{2}(\ket{10}+\ket{01})\otimes{(\ket{10}+\ket{01})}=\frac{1}{2}(\ket{10}\ket{10}+\ket{10}\ket{01}+\ket{01}\ket{10}+\ket{01}\ket{01})
% \end{equation}
% \begin{equation}
% \rightarrow \frac{1}{2}(\ket{10}\ket{00}+\ket{10}\ket{11}+\ket{01}\ket{11}+\ket{01}\ket{00})=\ket{\Psi^+}\ket{11}+...
% \end{equation}
% \begin{equation}
% \ket{\Psi^+}\ket{00}= \frac{1}{\sqrt{2}}(\ket{10}+\ket{01})\otimes(\ket{00}\rightarrow\frac{1}{\sqrt{2}}(\ket{10}\ket{10}+\ket{01}\ket{01})
% \end{equation}

\section{Comparison of Monte Carlo and analytical models} \label{AppendixA}
An analytical model to describe transducer-mediated conditional entanglement of remote qubits has been previously devised \cite{zeuthen2020figures}, where the output from two qubits would be converted to optical frequencies via transduction and subsequently interfere the signals on a beam splitter. A click in one of the detectors is an indication that the single-click scheme has succeeded. Subsequently applying a symmetric $\pi$-pulse to the qubits and conditioning on a second click in a 2-click scheme decreases the sensitivity to transduced noise photons. The focus of this appendix is to compare the analytical calculation of fidelity and compare it with our Monte Carlo simulation.

The analytical derivation starts with understanding the constituents of the remote entanglement procedure, which are the transduction + optical path efficiency ($\eta$) and the dark count rate ($p_d$). The transduction efficiency $\eta$ of the transducer can be modeled as:

\begin{equation}
    \eta = \eta_h \int_0^T |h_{out}(t)|^2 dt
\end{equation}

\noindent where $\eta_h = \int_{-\infty}^{\infty} \eta(\Omega) |h_{in}(\Omega)|^2 d\Omega$ is the mode-dependent efficiency and $h_{out}$ is the outgoing signal from the transducer in a time interval $T$ \cite{zeuthen2020figures}.  Each transducer contributes an average dark count rate of $\frac{r_N T}{2}$ in each detector, where $r_N$ is the added noise rate, whereby the probability for at least one dark count in a particular detector is:

\begin{align}
    p_d &= (1-e^{\frac{-r_N T}{2}})^2 = (1-e^{-r_N T}) \\
    \nonumber
    & \approx (1-1+r_N T + O(r_N T)^2) \approx r_N T
\end{align}

\noindent for $r_N T \ll 1$. It is important to note that the probability of having a dark count at a single detector is $r_N T$, but having a dark count in  either detector is $2 r_N T$. The resulting density matrix after heralding of a detected photon is $\rho_{1-click}' = \ket{\Psi} \bra{\Psi} $:

\begin{align} \label{one clicklick}
    \rho_{1-click}^{'} &= \frac{1}{N_1}((1-p_e)^2 2p_d(1-p_d)|00\rangle \langle 00| \\
    \nonumber
    & + 2p_e (1-p_e)\eta (1-p_d) |\Psi_+\rangle \langle\Psi_+| \\ \nonumber
    & + p_e (1-p_e)(1-\eta)(1-p_d)2p_d (|01\rangle \langle 01| + |10\rangle \langle 10|) \\ \nonumber
    & + p_e^2 ((1-(1-\eta)^2) + (1-\eta)^2 2p_d)(1-p_d)|11\rangle \langle11|
\end{align}

\noindent where $N_1$ is the normalization factor enforcing $Tr[\rho_{1-click}^{'}]=1$. $p_e$ represents the excitation probability of the superconducting qubits while $p_d$ represents the dark photon rate stemming from the noise rate given by the transducer and its efficiency $\eta$.
\renewcommand{\thefigure}{A\arabic{figure}}
\setcounter{figure}{0}
\begin{figure*}[h]
    \centering
    \begin{subfigure}[t]{0.5\textwidth}
        \centering
        \includegraphics[width = \linewidth]{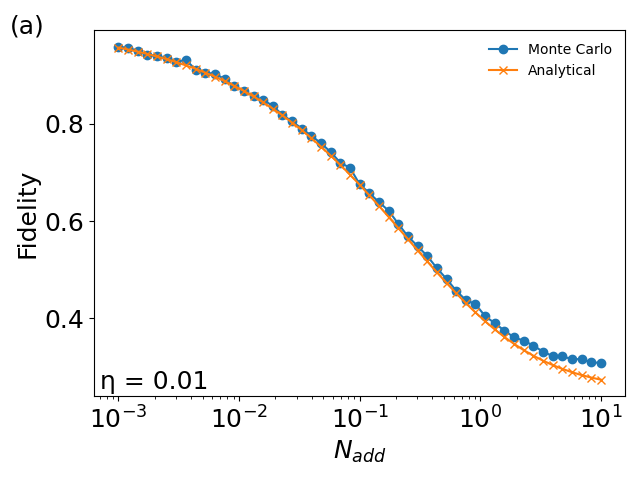}
    \end{subfigure}%
    \begin{subfigure}[t]{0.5\textwidth}
        \centering
        \includegraphics[width=\linewidth]{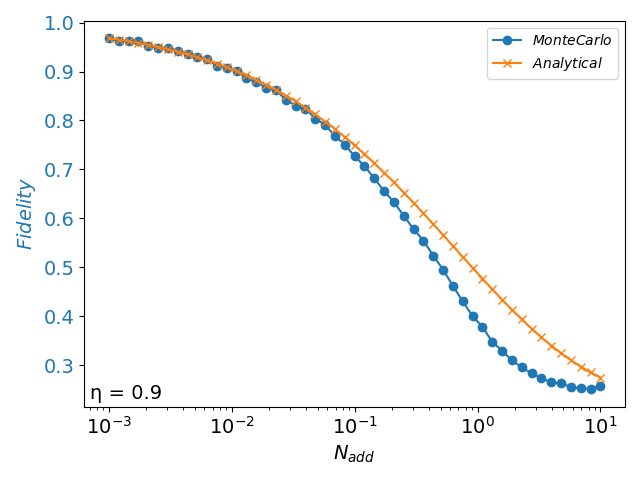}
    \end{subfigure}
    \caption{\justifying Fidelity for corresponding added noise $N_{add}$ at maximized excitation value $p_e$ for analytical \cite{zeuthen2020figures} and numerical results. Left shows the evolution of the fidelity for a transducer efficiency $\eta = 0.01$, the analytical and numerical agree almost perfectly wth the exception at larger $N_{add}$, where the analytical solutions starts to decay quickly to 0.25. Right Same results for $\eta = 0.9$ where the results still agree well with small deviations.}
    \label{fig:FideAnavsNum}
\end{figure*}

An analytical result for the fidelity of the single click protocol can then be derived by plugging equation \ref{one clicklick} into equation \ref{Fidelity}. Under the assumption that the conditions in the limit of $\frac{p_d}{\eta} \ll p_e \ll 1$, we obtain:

\begin{equation} \label{Fide1c}
    F_{1c} = \frac{2p_e (1-p_e)\eta + p_e (1-p_e) (1-\eta)2p_d}{p_e \eta (1-2p_d)(2-p_e \eta) + 2p_d}
\end{equation}

From \ref{Fide1c}, we compare the analytical and numerical results in \autoref{fig:FideAnavsNum} for $\eta = 0.01$ and $\eta = 0.9$. In both cases, $p_e$ is optimized for fidelity. We observe that the analytical result \cite{zeuthen2020figures} and numerical results agree well when $\eta = 0.01$ and $N_{add} < 1$. The divergences in the high $\eta$ / high $N_{add}$ regimes are due to discrepancies in the treatment of multiphoton events. Specifically, the analytical model specifically excludes analytical results, whereas the Monte Carlo simulation considers them. 
The 2-click scheme follows similar definitions as the 1-click scheme. The scheme works in two steps, and we will take as a condition that at least 1-click in exactly one arm occurs in each of the two steps. This results in $\rho_{2-click}' = \ket{\Psi} \bra{\Psi} $, the density matrix after the two-click protocol, we obtain: 

\begin{align}\label{two clicklick}
    \rho_{2-click}^{'} &= \frac{1}{N_2}((1-p_e)^2 2p_d(1-p_d) ([1-(1-\eta)^2](1-p_d)+(1-\eta)^2 2p_d (1-p_d))|00\rangle \langle 00| \\
    \nonumber
    & + 2p_e (1-p_e) \eta^2 (1-p_d)^2 |\Psi_+\rangle \langle\Psi_+| \\ \nonumber
    & +((\eta (1-p_d) + (1-\eta)(1-p_d)2p_d)^2 - \eta^2 (1-p_d)^2) (|01\rangle \langle 01| + |10\rangle \langle 10|) \\ \nonumber
    & + p_e^2 ((1-(1-\eta)^2) + (1-\eta)^2 2p_d)(1-p_d)^2 2p_d|11\rangle \langle11|)
\end{align}

From the equations \ref{one clicklick} and \ref{two clicklick} it is clear that the two-photon scheme has a smaller sensitivity to added noise than the one-photon scheme. On the other hand, the two-photon scheme will have a lower success probability if the transducer has a low efficiency since it requires the detection of two photons. 

\section{State of the Art Performance in Current Experimental Setups} \label{AppendixB}

\begin{table}[h]
\centering
\setcounter{table}{0}
\renewcommand{\arraystretch}{1.5} % Adjust row height for better spacing
\renewcommand{\thetable}{A\arabic{table}}
\setlength{\tabcolsep}{2pt} % Reduce column spacing

\begin{tabular}{ |c|p{2.0cm}|p{2.0cm}|p{3cm}|p{3cm}|p{3cm}|p{2.0cm}| } 
\hline
 & \makecell{\textbf{Bulk } \\ \textbf{LiNbO}$_3$ \cite{sahu2022quantum}}  
 & \makecell{\textbf{Thin-Film} \\ \textbf{LiNbO$_3$} \cite{warner2023coherent}}  
 & \makecell{\textbf{SiN} \\ \textbf{membrane} \cite{brubaker2022optomechanical}}  
 & \makecell{\textbf{Si/ LiNbO$_3$}\\ \textbf{P-O-M} \cite{weaver2024integrated}}  
 & \makecell{\textbf{Si / LiNbO$_3$}\\ \textbf{P-O-M} \cite{jiang2020efficient}}  
 & \makecell{\textbf{Si} \\ \textbf{O-M} \cite{zhao2024quantum}}  \\    
\hline
$\boldsymbol{\eta}$ & 8.7\% & 0.9\% & 47\% & $5.2\times10^{-5}$ & 5\% & 0.47\% \\ 
\hline
$N_{add}$ & 0.16 & 0.12 & 3.2 & 6 & 5 & 0.58 \\ 
\hline
\textbf{$f_{\text{rep}}$ (kHz)} & 0.5 & CW & CW & 100 & 170 & CW \\ 
\hline
\textbf{BW (MHz)} & 10 & 30 & 0.012 & 14.8 & 1.5 & 18 \\ 
\hline
\end{tabular}
\caption{Device performance parameters of several state-of-the-art transduction platforms.}
\label{tab:comparison}
\end{table}

\autoref{tab:comparison} shows the performance parameters of some of this generation's top-performing quantum transducers. The following key parameters are included: transducer efficiency ($\eta$), added noise ($N_{add}$), operating frequency ($f_{rep}$), and bandwidth (BW). The platforms considered include LiNbO$_3$ electro-optics,  LiNbO$_3$ piezo-optomechanics (P-O-M), and SiN membranes. In this table, $\eta$ is the microwave waveguide to optical waveguide efficiency, not the end-to-end optical + transduction efficiency as in the rest of the paper. The latter includes not only transduction efficiency, but also fiber coupling efficiency, optical filter efficiencies, and detector efficiencies. 

\section{Chi-Protocol Circuit Diagram} \label{AppendixC}
\renewcommand{\thefigure}{A\arabic{figure}}
\setcounter{figure}{1}
\begin{figure}[h]
    \centering
    \includegraphics[width=0.5\linewidth]{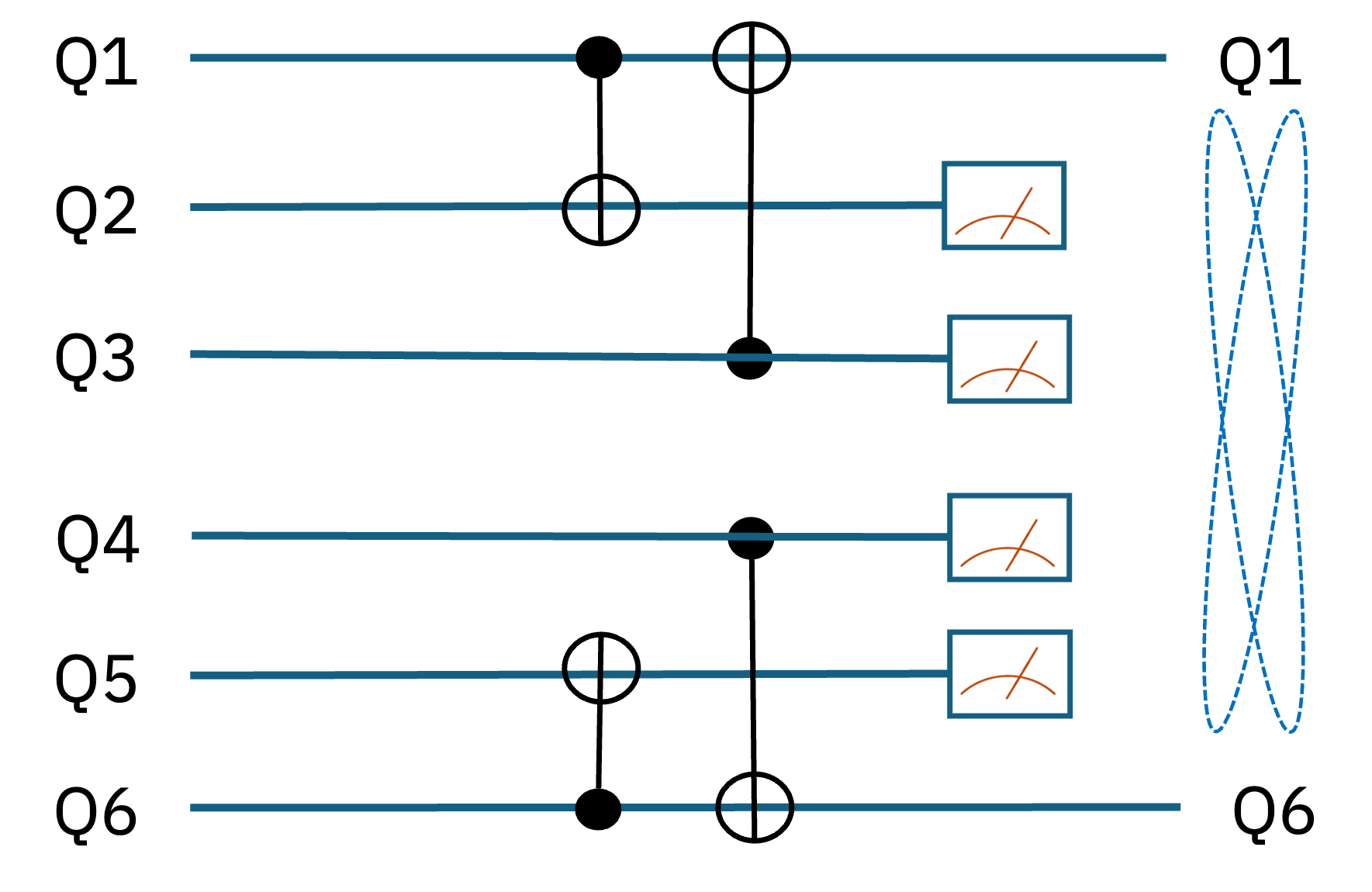}
    \caption{Chi protocol with local unitary operations $U_A$ and $U_B$ \cite{chi2012efficient}}
    \label{fig:Chi}
\end{figure}

\autoref{fig:Chi} displays the circuit diagram of the Chi protocol. This entanglement purification protocol works on three copies of a state via two bilateral controlled-NOT operations. The one-round success probability of the  protocol is larger than 2-to-1 protocols, but comes with the added constraint of having to prepare three different copies of Bell pairs.

\section{$T_1$ and $T_{2 \phi}$ degradation of Memory Qubit Fidelity}
In entanglement distillation, after the first Bell pair is heralded and stored in the memory, the memory can decay before the second Bell pair is heralded. We model this decay terms as a relaxation ($T_1$) process along with a dephasing ($T_{2 \phi}$) process using Kraus operator representations. 

\subsection{$T_1$ decay of the memory}
For $T_1$, we model and derive the mapping from the generalized amplitude-damping channel, which is a schematic model of the decay of an excited state of an atom due
to spontaneous emission of a photon or the spontaneous excitement of an atom due its environment. The Kraus operators for the generalized amplitude damping channel are given as:

\begin{align}
    M_0 &= \sqrt{p}\begin{pmatrix}
        1 & 0\\
        0 & \sqrt{1-\gamma}
    \end{pmatrix}
    \\
    M_1 &= \sqrt{p}\begin{pmatrix}
        0 & \sqrt{\gamma} \\
        0 & 0
    \end{pmatrix}
    \\
    M_2 &= \sqrt{1-p}\begin{pmatrix}
        \sqrt{1-\gamma} & 0 \\
        0 & 1
    \end{pmatrix}
    \\
    M_3 &= \sqrt{1-p}\begin{pmatrix}
        0 & 0 \\
        \sqrt{\gamma} & 0
    \end{pmatrix}
\end{align}

In the Kraus operators, $\gamma$ represents the probability of decay while $p$ represents the probability of excitement by the environment. The density matrix of the system after applyijng the Kraus map then evolves as:

\begin{align}
    S(\rho) = \begin{pmatrix}
        2\rho_{00} (1-\gamma) + p\gamma (\rho_{00} + \rho_{11}) & \sqrt{1-\gamma} \rho_{01} \\
        \sqrt{1-\gamma} \rho_{10} & \rho_{11} + \rho_{00} \gamma -\gamma p (\rho_{00} + \rho_{11})
    \end{pmatrix}
\end{align}

The probability of decay is related to $T_1$ as $\gamma = 1 - e^{-\frac{t}{T_1}}$. When looking at the state fidelity after applying the channel we obtain for a generic qubit state $\ket{\psi} = \alpha \ket{0} + \beta \ket{1}$:

\begin{equation}
    F = \bra{\psi}|\epsilon(\ket{\psi} \bra{\psi})|\ket{\psi} = \rho_{00}^2 (2 e^{-\frac{t}{T_1}} + p(1-e^{-\frac{t}{T_1}}) + \rho_{11}^2 (1-p(1-e^{-\frac{t}{T_1}})) + \rho_{01}\rho_{10} (2e^{-\frac{t}{2T_1}} + 1- e^{-\frac{t}{T_1}})
\end{equation}

% If $t\rightarrow \infty$ and $p=1$, we obtain:

% \begin{equation}
%     F = \rho_{00}^2 + \rho_{01}\rho_{10}
% \end{equation}

\subsection{$T_{2\phi}$ via Dephasing Channel}
The dephasing channel describes the interaction with the environment that can lead to loss of quantum information changes without any changes in qubit excitations. This can be modeled by the phase damping channel, with the following Kraus matrices:

\begin{align}
    M_0 &= \begin{pmatrix}
        1 & 0\\
        0 & \sqrt{1-\gamma}
    \end{pmatrix}
    \\
    M_1 &= \begin{pmatrix}
        0 & 0 \\
        0 & \sqrt{\gamma}
    \end{pmatrix}
\end{align}

The phase damping probability $\gamma \in [0,1]$ determines the amount of information lost to the environment. The subsequent state evolution is then:

\begin{equation}
    S(\rho) = \sum_i M_i \rho M_i^\dagger = \begin{pmatrix}
        \rho_{00} & (1-\gamma) \rho_{01} \\
        (1-\gamma) \rho_{10} & \rho_{11}
    \end{pmatrix}
\end{equation}

The  probability of phase damping is related to $T_{2_\phi}$ as:
$\gamma \approx \frac{1-e^{-\frac{t}{T_2}}}{2}$. The fidelity of the state after applying the channel can then be obtained as:

\begin{equation}
    F(T_{2_{\phi}}) = \frac{1+e^{-\frac{t}{T_2}}}{2} + \frac{1-e^{-\frac{t}{T_2}}}{2} (\rho_{00} - \rho_{11})
\end{equation}

If $t\rightarrow \infty$ then the fidelity becomes:

\begin{equation}
    F = \frac{1}{2} (1+(\rho_{00} - \rho_{11}))
\end{equation}

which can be observed in \autoref{fig:enter-label} (b) where for $T_2 \ll 1$ the fidelity tends to 0.5. 

\renewcommand{\thefigure}{A\arabic{figure}}
\setcounter{figure}{2}
\begin{figure}[h]
    \centering
    \includegraphics[width=0.5\linewidth]{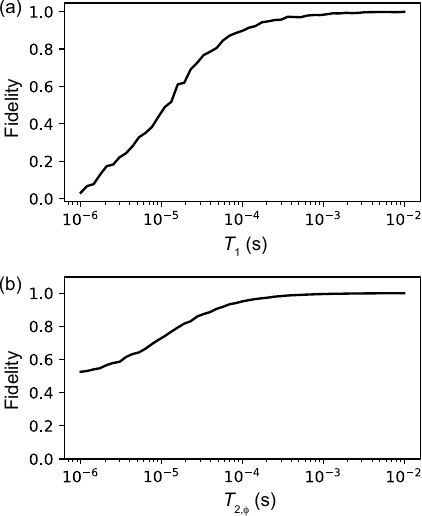}
    \caption{Dependence of distilled fidelity on $T_1$ and $T_{2,\phi}$}
    \label{fig:enter-label}
\end{figure}

\section{Local phase stability in the EPL Protocol}

After the remote entanglement procedure with which an entangled pair is created, often distillation procedures are necessary to filter out unwanted excitations or losses. The EPL protocol allows for the correction of photon loss and stabilization of unwanted local phases that occur during the entanglement procedure as a parity check. In this section, we analyze the application of the EPL protocol to our system where a dual-rail distillation procedure is applied on the generated entangled pairs. The pairs can be divided into four configuration subspaces coming from the remote entanglement procedure:

\begin{equation}
\begin{aligned}
H_0 &= \mathrm{span}(|gg\rangle),\\
H_1 &= \mathrm{span}(|\Psi^\pm_\phi \rangle), \\
H_2 &= \mathrm{span}(|ge\rangle
\langle ge|+ |eg\rangle
\langle eg|),\\
H_3 &= \mathrm{span}(|ee\rangle),
\end{aligned}
\end{equation}

\noindent and write the single-copy state as

\begin{equation}
\rho = N\,\Pi_{H1,\Psi_\phi} + K\,\Pi_{H2} + L\,\Pi_{0} + M\,\Pi_{3},
\qquad N + K + L + M = 1,
\end{equation}

where

\begin{equation}
\Pi_{H1,\Psi_\phi}=|\Psi_\phi\rangle\!\langle\Psi_\phi|,
\quad
|\Psi^\pm_\phi\rangle=\frac{1}{\sqrt{2}}\!\left(|ge\rangle \pm e^{i\phi}|eg\rangle\right),
\quad
\Pi_{g}=|gg\rangle\!\langle gg|,
\quad
\Pi_{e}=|ee\rangle\!\langle ee|.
\end{equation}

\noindent We take two identical copies:
\begin{equation}
\rho_{\text{start}}=\rho \otimes \rho .
\end{equation}

These copies are subject to bilateral CNOTs with one pair as control and the second as target,

\begin{equation}
U=\mathrm{CNOT}_{A_1\to A_2}\otimes \mathrm{CNOT}_{B_1\to B_2}.
\end{equation}

After the bilateral CNOT's, the state is measured and only state $|ee\rangle$ is post selected. Algebraically the whole primitive is the completely positive (CP) map

\begin{equation}
\mathcal{E}(\cdot)=K_{\mathrm{EPL}}(\cdot)K_{\mathrm{EPL}}^\dagger,\qquad
K_{\mathrm{EPL}}=|ge\rangle\!\langle ge|\otimes \langle eg|+|eg\rangle\!\langle eg| \otimes \langle ge|.
\end{equation}

$K_{\mathrm{EPL}}$ is the “both nodes disagree across the two copies’’ parity filter written as a Kraus operator. $K_{EPL}$ is the operator that takes a state on two copies ($\rho_1 \otimes \rho_2$) and returns a state on one copy. It has support only on $H_1^{\otimes 2}$; it annihilates $H_0$, $H_3$, and anything orthogonal to $|ge\rangle,|eg\rangle$ (the whole $H_2$). The unnormalized post-selected control state is

\begin{equation}
\sigma=K_{\mathrm{EPL}}(\rho\otimes\rho)K_{\mathrm{EPL}}^\dagger,\qquad
p_{\mathrm{succ}}=\mathrm{Tr}\,\sigma,\qquad
\rho_{\mathrm{out}}=\sigma/p_{\mathrm{succ}}.
\end{equation}

Looking at the local phase $\phi$ that occurs during differences in the optical path lenght of the photons, we can model the slowly drifting local phase as a unitary

\begin{equation} 
U_\phi|ge\rangle=|ge\rangle,\qquad
U_\phi|eg\rangle=e^{i\phi}|eg\rangle,\qquad
U_\phi= I \ \text{on } H_0\oplus H_2\oplus H_3.
\end{equation}

Assuming that both copies will share the same $\phi$, we apply the Kraus formalism on the two copies:

\begin{equation}
\rho\otimes\rho \ \mapsto\ (U_\phi\!\otimes\!U_\phi)(\rho\otimes\rho)(U_\phi^\dagger\!\otimes\!U_\phi^\dagger).
\end{equation}

Only the $|ge,eg\rangle$ and $|eg,ge\rangle$ terms pass the filter thus leading to,

\begin{equation}
K_{\mathrm{EPL}}(U_\phi\!\otimes\!U_\phi)=e^{i\phi}K_{\mathrm{EPL}},
\end{equation}

Hence, the local phase after applying the map representative of the EPL protocol becomes global:

\begin{equation}
K_{\mathrm{EPL}}(U_\phi\!\otimes\!U_\phi)(\rho\otimes\rho)(U_\phi^\dagger\!\otimes\!U_\phi^\dagger)K_{\mathrm{EPL}}^\dagger
=K_{\mathrm{EPL}}(\rho\otimes\rho)K_{\mathrm{EPL}}^\dagger,
\end{equation}

for any $\rho$. Therefore the EPL output is independent of $\phi$: the common $e^{i\phi}$ becomes a global phase and disappears.

\section{Imperfect Gates and Measurements}

We model a noisy CNOT gate as:

\begin{equation}
    \overline{U} = N_{gate} \circ U_{CNOT}
\end{equation} 

\noindent where $N_{gate}$ is a decay channel characterizing the faulty CNOT gate. Figure \ref{fig:heatmap} shows the effect of CNOT gate errors on transduction fidelity using a depolarizing channel to mimic the noise on the CNOT operation. The depolarization channel is defined via its Kraus matrices as:

\begin{align}
    K_0 &= \sqrt{1-p} \begin{pmatrix}
        1 & 0 \\
        0 & 1
    \end{pmatrix} \\
    K_1 &= \sqrt{\frac{p}{3}} \begin{pmatrix}
        0 & 1 \\
        1 & 0
    \end{pmatrix} \\
    K_2 &= \sqrt{\frac{p}{3}} \begin{pmatrix}
        0 & -i \\
        i & 0
    \end{pmatrix} \\
    K_3 &= \sqrt{\frac{p}{3}} \begin{pmatrix}
        1 & 0 \\
        0 & -1
    \end{pmatrix}
\end{align}

\renewcommand{\thefigure}{A\arabic{figure}}
\setcounter{figure}{3}

\noindent where $p\in \{0,1\}$ is the depolarization probability and is equally divided in the application of all Pauli operations. At $p=0$, the channel will be a noise-free channel. For $p=\frac{3}{4}$, the channel will be a fully depolarizing channel. At $p=1$, the channel will be a uniform Pauli error channel. In the simulation, we define $\epsilon = 1-p$ as the noise parameter. The noise channel is then applied from the two pair input $\rho \otimes \rho $, where we apply the Kraus map of the faulty operators:

\begin{equation}
    (\Bar{U} \otimes \Bar{U}) \rightarrow \sum_{a,b} (K_a \otimes K_b) (\rho \otimes \rho) (K_a \otimes K_b)^\dagger
\end{equation}

\noindent where $K_a$ are the Kraus operators of $\Bar{U}$. 

\begin{figure}[h]
    \centering
    \includegraphics[width=0.6\linewidth]{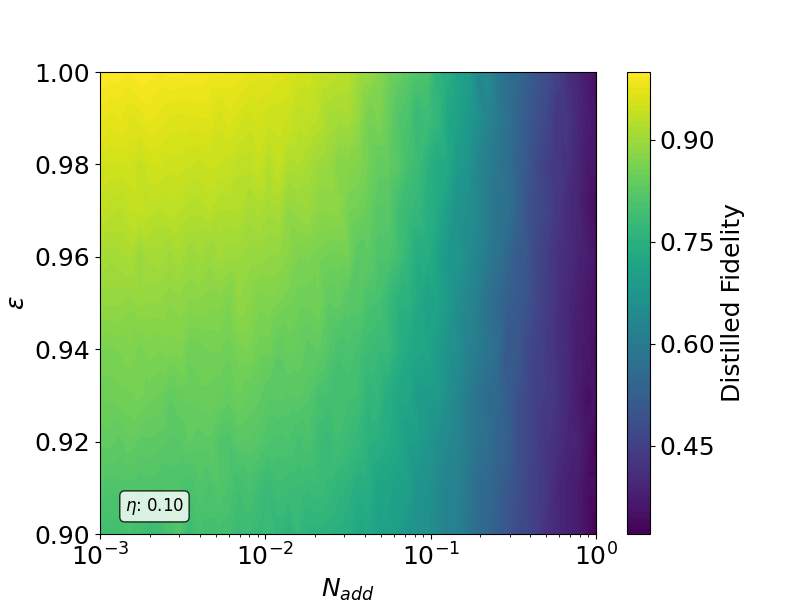}
    \caption{\justifying Distilled fidelity heatmap for imperfect CNOT gate as a function of $N_{add}$ and gate error $\epsilon$. $\epsilon = 1$ indicates a perfect CNOT gate. The distilled fidelity is computed for a fixed excitation probability of $p_e = 0.5$ and $\eta = 10\%$.}
    \label{fig:heatmap}
\end{figure}

\section{Noise model premise}
\label{AppendixNoiseModel}
In this appendix, we provide a heuristic reasoning underlying the noise model depicted in Table~\ref{table:monte} of the main text. 
Consider the four party state state $\ket{\Psi}_{A,B}$ in eq.~\eqref{eq:keyState}, where the microwave system is written using the basis $\{\ket{e}, \ket{g}\}$ and the fock basis is used to write the optical channels $c$ and $d$.
When these optical channels are post-selected for a single photon in $c$ or $d$ but not both, one obtains a state, up to normalization, of the form
\begin{equation}
\ket{\Psi_1}_{A,B} \propto \ket{\Psi^{+}} \ket{01}_{cd} + \ket{\Psi^{-}} \ket{10}_{cd},
\end{equation}
where $\ket{\Psi^\pm}=(\ket{eg}\pm\ket{ge})/\sqrt{2}$. If the target Bell state for the full transduction process is $\ket{\Psi^+}$ then this target is obtained when there is a click due to a photon in $d$, or by applying a phase flip $Z$~($Z := \ket{g}\bra{g} - \ket{e}\bra{e}$) when there is a click from $c$.
Let $\mathcal{P}$ refer to this procedure of obtaining this target Bell state.
The noiseless procedure $\mathcal{P}$ of obtaining the target Bell state becomes noisy when the optical photon undergoes a phase flip $\ket{i} \mapsto (-1)^i \ket{i}$ prior to detection. Due to this noise, the procedure $\mathcal{P}$ results in a microwave state $\ket{\Psi^{-}}$, which carries a $Z$ error with respect to the target Bell state $\ket{\Psi^{+}}$. Two additional sources of noise are loss or excitation of a photon in the optical channel. If there is a loss in $c$ or $d$, the resulting microwave state is 
$\ket{ee}$. However, if there is an excitation of a photon in $c$ or $d$ then the resulting microwave state is $\ket{gg}$. Together, these three noise processes result in a noisy microwave state given by a classical mixture of pure states
$\{\ket{\Psi^{+}}, \ket{gg}, \ket{ee}, \Psi^{-}\}$. This mixture can also be re-parametrized as a mixture of mixed states 
\begin{equation}
\{\ket{\Psi^{+}}\bra{\ket{\Psi^{+}}},
\ket{gg}\bra{gg}, \ket{ee}\bra{ee}, (\ket{\Psi^{+}}\bra{\ket{\Psi^{+}}} + \ket{\Psi^{-}}\bra{\ket{\Psi^{-}}})/2 \}.
\end{equation}
The first state, $\ket{\Psi^+}\bra{\Psi^+}$,
represents a noiseless target Bell state, the second $\ket{gg}\bra{gg}$, appears as an amplitude damped version of the target, the third, $\ket{ee}\bra{ee}$, appears as a generalized amplitude damped state of the target where excitation to $\ket{e}$ can occur
in addition to damping to $\ket{g}$, while the fourth state
$(\ket{\Psi^{+}}\bra{\ket{\Psi^{+}}} + \ket{\Psi^{-}}\bra{\ket{\Psi^{-}}})/2$ represents a loss of coherence in the two dimensional subspace of the Bell states
$\{ \ket{\Psi^{+}},  \ket{\Psi^{-}} \}$. 
\twocolumngrid

\bibliographystyle{naturemag} % Or any preferred style
\bibliography{references}

\end{document}